\documentclass[epjST]{svjour}

\usepackage{graphicx}
\graphicspath{ {graphics/} }
\usepackage[utf8]{inputenc}

\usepackage[breaklinks=true, unicode=true, colorlinks=true, linkcolor=black, citecolor=black, urlcolor=black]{hyperref}
\usepackage{doi}

\usepackage{amssymb}

\usepackage{siunitx}

\begin{document}

\title{High-harmonic spectra of hexagonal nanoribbons from real-space time-dependent Schrödinger calculations}
\author{Helena Drüeke\thanks{\email{helena.drueeke@uni-rostock.de}} \and Dieter Bauer\thanks{\email{dieter.bauer@uni-rostock.de}}}

\institute{University of Rostock, Germany}

\abstract{
High-harmonic spectroscopy is a promising candidate for imaging electronic structures and dynamics in condensed matter by all-optical means and with unprecedented temporal resolution.
We investigate harmonic spectra from finite, hexagonal nanoribbons, such as graphene and hexagonal boron nitride, in armchair and zig-zag configuration.
The symmetry of the system explains the existence and intensity of the emitted harmonics.
}

\maketitle

\section{Introduction}
\label{sec:introduction}
High-harmonic generation (HHG) has been first observed in gases~\cite{mcpherson_studies_1987,ferray_multiple-harmonic_1988}.
Its non-perturbative nature, featuring a plateau of almost constant high-harmonic yield, was subsequently explained by the three-step model~\cite{corkum_plasma_1993,lewenstein_theory_1994}:
An electron is removed from the atom, propagates under the external field's influence, and recombines with the atom.
The orbital energies of electrons in atoms do not depend on momentum, and the electron's dispersion relation in the continuum is shaped parabolically.
Therefore, no harmonics are emitted from electrons in the ground state or free electrons, only by transitions between bound states or recombination from continuum states back to bound states.

To describe HHG in the bulk of solids, the orbital energies and the continuum are replaced by electronic bands~\cite{vampa_merge_2017}.
This opens a whole new field of research~\cite{ghimire_observation_2011,ghimire_generation_2012,ghimire_strong-field_2014,hohenleutner_real-time_2015,ndabashimiye_solid-state_2016,ikemachi_trajectory_2017,you_high-harmonic_2017,yu_high-order_2019}.
Analogous to the HHG process in gases, the transition of electrons between valence and conduction bands causes high harmonics, called interband harmonics.
Intraband harmonics, on the other hand, are produced by the movement of electrons in partially filled, non-parabolic bands.
Band structures are usually defined for periodic or infinite solid bulk systems. However, every realistic system has boundaries, which may cause completely different HHG spectra compared to the bulk~\cite{bauer_high-harmonic_2018,drueke_robustness_2019,jurs_high-harmonic_2019,hansen_finite-system_2018,chacon_observing_2020}.
Graphene and hexagonal boron nitride (hBN) are two-dimensional materials that possess  fascinating features with promising potential applications~\cite{castro_neto_electronic_2009,wang_graphene_2017}.
Their hexagonal structure allows for two different edges: zig-zag and armchair.
While graphene consist only of carbon atoms (all identical), hBN is built from boron and nitrogen atoms.
Recently, the interaction of intense laser light with graphene got into the focus of interest for its prospects to steer electrons at will on ultrafast time scales~\cite{koochaki_kelardeh_graphene_2017,higuchi_light-field-driven_2017,heide_coherent_2018,baudisch_ultrafast_2018}.

\section{Methods}
\label{sec:methods}
The nanoribbons' atomic nuclei were positioned in a hexagonal lattice, as described for the armchair and zig-zag configuration in the following sections \ref{sec:armchair} and \ref{sec:zig-zag}.
Atomic units are used throughout this paper unless stated otherwise.
The distance between neighboring lattice sites was $2.683$, the bond length in graphene ($\SI{1.42}{\angstrom}$).
An effective Pöschl-Teller potential
$
V(\vec{r}) = - \sum_i \frac{V_i}{\cosh^2(\varepsilon |\vec{r} - \vec{r}_i|)}
$
with ion potentials $V_i = 3.2 \pm V_\mathrm{os}$
and screening parameter $\varepsilon = 2$ describe the attractive potentials of the nuclei.
For graphene ribbons, all atoms are carbon, therefore the additional on-site potential $V_\mathrm{os} = 0$.
A non-zero on-site potential represents two alternating, different kind of atoms, such as boron and nitrogen in hBN.
At which lattice sites the ion potentials are increased or decreased by $V_\mathrm{os}$ is sketched in the following sections.

In this work, we did not employ the usual tight-binding approximation commonly made in condensed-matter theory but have developed a 2D, real-space, time-dependent Schrödinger solver for the {\em ab initio} simulation of the intense-laser interaction with 2D matter. In that way we are able to reveal differences and similarities in HHG spectra as compared to corresponding tight-binding studies, e.g., in Ref.~\cite{jurs_high-order_2021}.
The non-interacting electronic orbitals in our Schrödinger solver are defined on a two-dimensional grid of spacing $\Delta x = \Delta y = 0.2$, which encompasses all lattice sites plus a border of $8$ on each side.
In contrast to the usual tight-binding description, this allows us to have electron orbitals that are not only localized at lattice sites but also between them, or free electrons.

The electronic eigenstates of the system were found by imaginary-time propagation employing the Crank-Nicolson method~\cite{bauer_ed._computational_2017}.
Starting from a random initialization, imaginary timesteps $-0.05\mathrm{i}$ are taken (each step followed by renormalization of the wavefunction) until the ground state is reached and the relative change of the state is smaller than the threshold of $10^{-18}$ for two consecutive iterations.
To find the higher-lying states, the workflow is identical, but with an additional (Gram-Schmidt) orthogonalization to all previously found states in each iteration.
This gives us all states of interest of the unperturbed system.

Real-time simulations of the interaction of all occupied electronic orbitals with a short laser pulse were performed with a timestep $0.05$ using, again, Crank-Nicolson propagation.
The pulse was a 4-cycle $\sin^2$-shaped laser pulse of frequency $\omega = 0.0075$ ($\lambda \simeq \SI{6.1}{\micro\meter}$) and polarized along the ribbon.
The electronic dipoles were recorded at each time step during the laser pulse.
Harmonic spectra were calculated as the absolute square of the Fourier transform of the recorded dipoles, multiplied by a symmetric Hann window~\cite{baggesen_dipole_2011,harris_use_1978}.

\section{Armchair ribbon}
\label{sec:armchair}
\begin{figure}[htbp]
\centering
\includegraphics{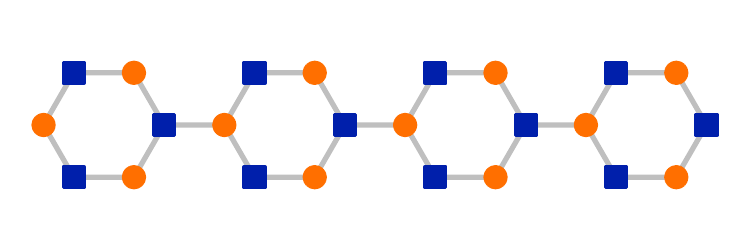}
\caption{
Armchair ribbon:
At the blue (square) lattice sites, the potential depth is increased by the on-site potential, $V_\mathrm{blue} = V_\mathrm{avg} + V_\mathrm{os}$.
The orange (circle) sites correspond to the shallower potentials of depth $V_\mathrm{orange} = V_\mathrm{avg} - V_\mathrm{os}$.
}
\label{fig:armchair}
\end{figure}

First, we investigate a hexagonal nanoribbon in the armchair configuration.
A total of 24 lattice sites are arranged in the shape of four hexagons  as shown in \autoref{fig:armchair}.
For $V_\mathrm{os} = 0$, the armchair ribbon is symmetric about the horizontal as well as the vertical axis through the center.
The introduction of an on-site potential deepens the blue (square) sites' potentials while making the orange (circle) ones shallower.
This causes a left-right asymmetry, while the top-bottom symmetry is conserved.
Note that the lines drawn in \autoref{fig:armchair} connect nearest neighbors.
In tight-binding calculations (such as in Ref.~\cite{jurs_high-order_2021}), hopping takes place along these lines.
However, in our simulation based on the time-dependent Schrödinger equation, electronic wavefunctions are not restricted to move along these lines but may propagate in the entire plane.

\begin{figure}[htbp]
\centering
\includegraphics{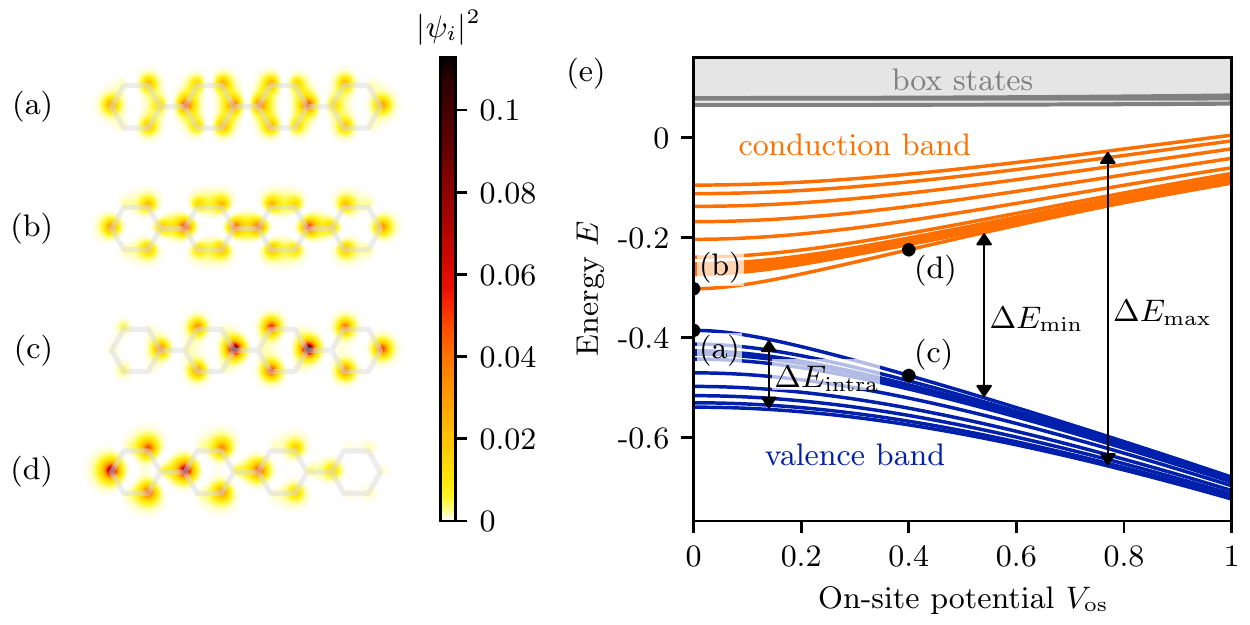}
\caption{
Orbitals and orbital energies of the armchair ribbon.
(a) - (d) Orbitals of the armchair ribbon.
(a)~Highest occupied orbital without on-site potential ($V_\mathrm{os} = 0$).
(b)~Lowest unoccupied orbital without on-site potential ($V_\mathrm{os} = 0$).
(c)~Highest occupied orbital with on-site potential $V_\mathrm{os} = 0.4$.
(d)~Lowest unoccupied orbital with on-site potential $V_\mathrm{os} = 0.4$.
(e)~Orbital energies of the armchair ribbon as a function of on-site potential $V_\mathrm{os}$.
The orbital energies of the orbitals shown in (a) - (d) are marked.
The arrows mark the bandwidth of the valence band ($\Delta E_\mathrm{intra}$) and the minimum and maximum bandgap between the valence and (first) conduction band ($\Delta E_\mathrm{min}$ and $\Delta E_\mathrm{max}$).
}
\label{fig:armchair_orbitals_energies}
\end{figure}

The asymmetry in the potential leads to an asymmetry in the orbitals.
\autoref{fig:armchair_orbitals_energies}~(a-d) show the highest occupied
(a and c) and lowest unoccupied (b and d) orbitals without (a and b) and with (c and d) on-site potential.
The orbitals without on-site potential are horizontally and vertically symmetric (as is the potential), and there is only a small bandgap between the occupied and unoccupied states.
With an on-site potential, the occupied orbitals are localized on the sites with deeper potentials and therefore have a decreased energy.
The unoccupied orbitals are localized on the sites with shallower potentials and therefore have increased energy.
This leads to a bandgap between the occupied and unoccupied orbitals, which, for $V_\mathrm{os} \gtrsim 0.2$, grows linearly with the on-site potential (\autoref{fig:armchair_orbitals_energies}~(e)).
In contrast to tight-binding methods, our approach allows us to calculate an arbitrary number of orbitals of increasing energy.
The next state above the conduction band is a "free" electron, i.e., not localized on the ribbon but still inside the simulation box with reflecting boundary conditions. 

\begin{figure}[htbp]
\centering
\includegraphics{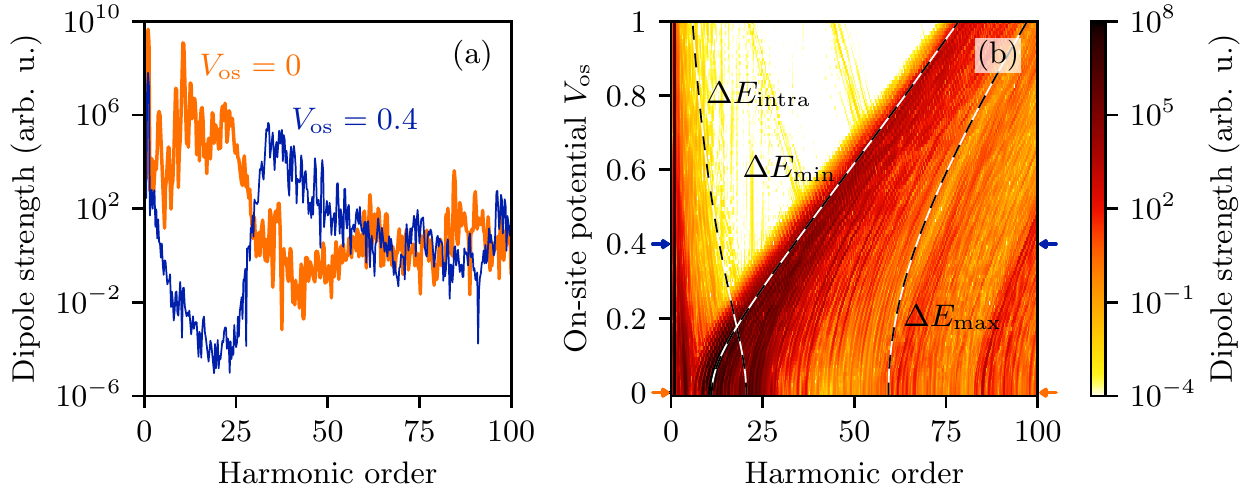}
\caption{
Harmonic spectra of the armchair ribbon in parallel direction.
(a)~Harmonic spectra without (bold orange line, $V_\mathrm{os} = 0$) and with (blue line, $V_\mathrm{os} = 0.4$) on-site potential.
(b)~Harmonic spectra as a function of harmonic order and on-site potential.
}
\label{fig:armchair_harmonics}
\end{figure}

The incoming laser field is linearly polarized along the armchair ribbon.
All emitted harmonics are linearly polarized in the same direction.
The emission of harmonics polarized in the perpendicular direction requires a top-bottom asymmetry in the system, which the armchair ribbon does not possess, regardless of on-site potential.
The bandwidths and bandgaps ($\Delta E_\mathrm{intra}$, $\Delta E_\mathrm{min}$, and $\Delta E_\mathrm{max}$) from \autoref{fig:armchair_orbitals_energies}~(e) explain the most important features of the harmonic spectra shown in \autoref{fig:armchair_harmonics}.
Intraband harmonics are only present at harmonic energies below the width of the valence band $\Delta E_\mathrm{intra}$.
Interband harmonics can be observed between the minimum $\Delta E_\mathrm{min}$ and maximum $\Delta E_\mathrm{max}$ bandgap between the valence and (first) conduction band.
Above the two bands are the box states (marked as gray lines in \autoref{fig:armchair_orbitals_energies}~(e)), which are not localized on the ribbon, and whose energies are determined by the size of the simulation box.
Only the energies of the four lowest box states are shown, but many more lie above them.
Transitions to these box states cause harmonics above $\Delta E_\mathrm{max}$.
In an experiment, there are no box states (unless it is performed in a cavity), but transitions to higher bands or the continuum would also cause harmonics beyond the maximum bandgap.
These can not be described in tight-binding approximation with one atomic orbital per site because then the energy difference between states is bound from above by $\Delta E_\mathrm{max}$ (see Ref.~\cite{jurs_high-order_2021}).

\section{Zig-zag ribbon}
\label{sec:zig-zag}

\begin{figure}[htbp]
\centering
\includegraphics{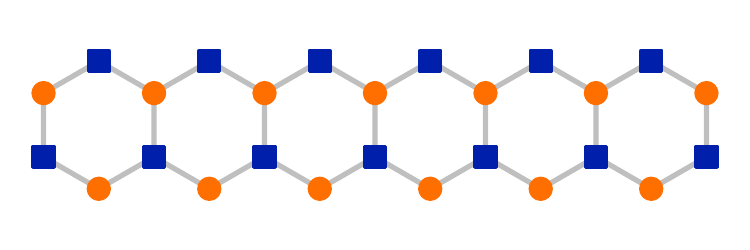}
\caption{
Zig-zag ribbon:
At the blue (square) lattice sites, the potential depth is increased by the on-site potential, $V_\mathrm{blue} = V_\mathrm{avg} + V_\mathrm{os}$.
The orange (circle) sites correspond to the shallower potentials of depth $V_\mathrm{orange} = V_\mathrm{avg} - V_\mathrm{os}$.}
\label{fig:zig-zag}
\end{figure}

In the zig-zag configuration, a total of 26 lattice sites are arranged in six hexagons, as shown in  \autoref{fig:zig-zag}.
On the orange sites, the on-site potential decreases the potential depth, while on the blue sites, it deepens the potential.
The on-site potential causes a top-bottom asymmetry but no left-right asymmetry.

\begin{figure}[htbp]
\centering
\includegraphics{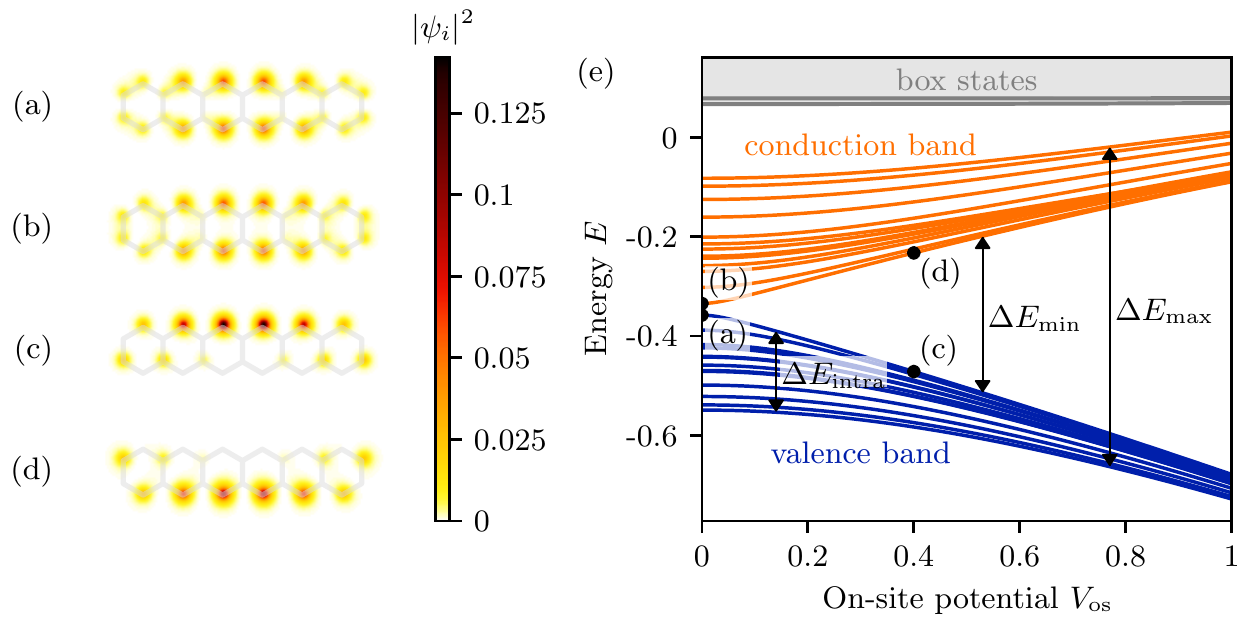}
\caption{
Orbitals and orbital energies of the zig-zag ribbon.
(a) - (d) Orbitals of the zig-zag ribbon.
(a)~Highest occupied orbital without on-site potential ($V_\mathrm{os} = 0$).
(b)~Lowest unoccupied orbital without on-site potential ($V_\mathrm{os} = 0$).
(c)~Highest occupied orbital with on-site potential $V_\mathrm{os} = 0.4$.
(d)~Lowest unoccupied orbital with on-site potential $V_\mathrm{os} = 0.4$.
(e)~Orbital energies of the zig-zag ribbon as a function of on-site potential $V_\mathrm{os}$.
The orbital energies of the orbitals shown in (a) - (d) are marked.
The arrows mark the bandwidth of the valence band ($\Delta E_\mathrm{intra}$) and the minimum and maximum bandgap between the valence and (first) conduction band ($\Delta E_\mathrm{min}$ and $\Delta E_\mathrm{max}$).
}
\label{fig:zig-zag_orbitals_energies}
\end{figure}

As for the armchair ribbon, the asymmetry of the potential leads to decreased energies of states in the valence band, localized at the deeper sites, and increased energies of states in the conduction bands, localized at the shallower sites (see \autoref{fig:zig-zag_orbitals_energies}).
The minimum bandgap increases almost linearly with the on-site potential, the bandwidths of both bands decrease.

\begin{figure}[htbp]
\centering
\includegraphics{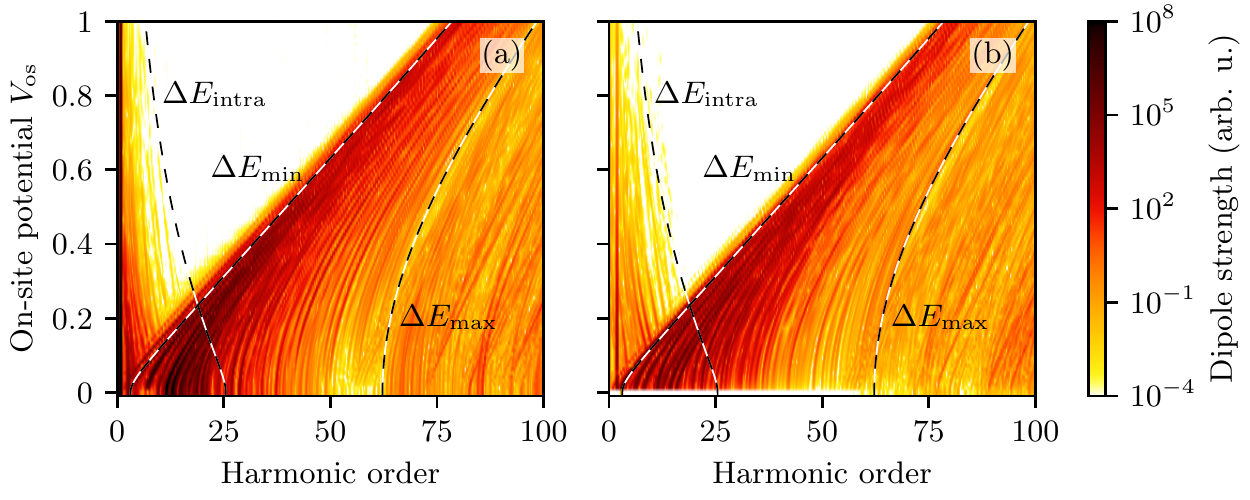}
\caption{
Harmonic spectra of the zig-zag ribbon as a function of on-site potential in
(a)~parallel and
(b)~perpendicular polarization.}
\label{fig:zig-zag_harmonics}
\end{figure}

The parallelly polarized harmonics (\autoref{fig:zig-zag_harmonics} (a)) are present with and without on-site potential.
The bandwidth and bandgaps can explain the cutoffs of both intra- and interband harmonics.
Perpendicular harmonics (\autoref{fig:zig-zag_harmonics} (b)) with on-site potential agree with these cutoffs, as well.
Without an on-site potential, there is no top-bottom asymmetry, and therefore almost no harmonics perpendicular to the laser are observed.
Transitions to the box states lead to weak harmonic emission above $\Delta E_\mathrm{max}$.

\section{Conclusion}
\label{sec:conclusion_outlook}
The introduction of an on-site potential in hexagonal nanoribbons causes lower energies for occupied states and higher energies for unoccupied states.
The valence band's bandwidth decreases, and the minimum and maximum bandgaps between the valence and conduction bands increase.
These three energies explain the overall features in harmonic spectra for different on-site potentials.
Intraband harmonics are only present at energies below the valence bandwidth.
Interband harmonics are present at energies between the minimum and maximum bandgap.
For a laser polarized along the ribbon, the resulting harmonics are polarized in the same direction unless a non-zero on-site potential causes a top-bottom asymmetry, which is only possible in the zig-zag ribbon.

The results of this paper provide valuable verification of simpler tight-binding models~\cite{jurs_high-order_2021}.
Our approach is not limited to a fixed number of states (grouped in bands), and our results account for transitions to even higher bands or the continuum. However, these transitions are expected to play an important role only in the generation of  higher harmonics beyond the cutoff $\Delta E_\mathrm{max}$, leading to higher-order plateaus with decreasing yield (see, e.g.,~\cite{hansen_high-order_2017}).
On the other hand, tight-binding approaches capture the essential mechanisms underlying high-harmonic generation up to $\Delta E_\mathrm{max}$, are computationally much less demanding and thus can be used to investigate much larger systems.

The datasets generated and analyzed during this study are available at\\
\doi{10.17605/OSF.IO/8RTFU}~\cite{drueke_high-harmonic_2021}.

\bibliography{Library}

\end{document}